\documentstyle[12pt]{article}
\begin{document}
\title{Bridging the Dimensional Gap: from  Kink in One Dimension to 
Curved Domain Wall in Three Dimensions\thanks{Presented at XXXVIII 
Cracow School of Theoretical Physics, Zakopane, 1-10 June 1998.}}
\author{H. Arod\'z \\ 
Institute of Physics, Jagellonian University
\\  Reymonta 4, 30-059 Cracow, Poland.}
\maketitle
\vspace*{1cm}
\begin{abstract}
Improved expansion in width is applied to a curved domain wall in 
nonrelativistic dissipative $\lambda(\Phi^2 - v^2)^2 $ model
with real scalar order parameter $\Phi$. Approximate analytic description 
of such a domain wall to second order in the width is presented.
\end{abstract}

\vspace*{2cm}

\noindent Preprint TPJU-25/98 
\pagebreak
  
\section{Introduction }
Physics of domain walls and vortices has been a rather interesting field of 
experimental as well as theoretical research for quite a long time \cite{1}.
Example of a recent hot topic is production of such soft solitonic 
objects in rapid phase transitions --- 
testing and refining a theoretical description proposed 
by Kibble \cite{2} and \.Zurek \cite{3}.  Theoretical analysis of dynamics of
domain walls and vortices is relatively difficult because pertinent
field equations are nonlinear, and the most interesting solutions do not 
belong to weak field sector.  

Among problems which have been discussed in literature is  time evolution
of single curved domain wall or vortex. It is commonly regarded as 
accessible only by a numerical analysis. Actually, there exist also analytical
approaches which yield (an approximate) description of the time evolution:
the classical effective action  (CEA) method which has been developed in a 
series of papers starting from \cite{4}, \cite{5}, examples of more recent 
works are \cite{6}, \cite{7}, \cite{8}, and a version of Hilbert --- Chapman
--- Enskog method which we call the improved expansion in  width
(IEW). This latter approach has been developed in papers \cite{9}, with
an inspiration coming from \cite{10}. The two methods have been outlined and
compared in  paper \cite{11}. These two analytical approaches are not
simple, but neither is the purely numerical approach --- this is just a 
reflection of the fact that dynamics of the curved domain wall or
a vortex is nontrivial due to nonlinearity, many spatial dimensions, and
many modes of involved fields. The numerical and analytical approaches
should be regarded as equally important and complementary sources of 
information about the dynamics. 

In the present paper we apply IEW method to a curved domain wall in a 
nonrelativistic dissipative system.   
Time evolution is governed by a diffusion type equation for which no 
simple action functional exists. Therefore, it is not clear how  CEA scheme
could be applied in this case, while, as it turns out,  IEW method works 
quite well. Because our second goal is a presentation of the method, 
we consider a relatively simple system with scalar order parameter. 
An application to domain walls in nematic liquid crystals we will present
elsewhere \cite{12}. IEW method has also been applied to a vortex line, 
see \cite{13}. Work on application to a disclination line in a nematic 
liquid crystal is in progress \cite{14}.

In our opinion IEW method has several attractive features, e.g., it 
combines the old and elegant subject of differential geometry of surfaces in
3-dimensional space with nonlinear dynamics of the curved domain wall.
Another interesting aspect is that IEW scheme relates properties of the 
domain wall to properties of one-dimensional  kink. In this sense IEW
method embodies the idea that the curved domain wall can be regarded as 
three-dimensional embedding of the one-dimensional kink. 

The expansion in  width is based on the idea that transverse profile of 
the curved domain wall considered in suitable coordinates  (which
are called comoving coordinates) differs from transverse profile of 
a planar domain wall by small corrections which are due to curvature of 
the domain wall, and that these corrections can be calculated perturbatively. 
There is a condition for applicability of such a perturbative scheme:
the two main curvature radia of the domain wall should be much larger than 
its width.  As we shall see below, turning that idea into a concrete 
calculational scheme requires some work, but that should be expected in any
approach which tackles  generic curved domain walls. 

The plan of our paper is as follows. In Section 2 we introduce the comoving
coordinate system. Section 3 is devoted to the presentation of the 
perturbative expansion.  Several remarks are collected in Section 4.

\section{ The comoving coordinates }
We shall seek  the curved domain wall solutions of the following equation
\begin{equation}
\gamma \frac{\partial \Phi}{\partial t} + \frac{\delta F}{\delta \Phi} = 0,
\end{equation}
where the free energy $F$ has the form
\begin{equation}
F = \frac{1}{2}\int d^3x \left(\frac{\partial\Phi}{\partial x^{\alpha}}  
\frac{\partial\Phi}{\partial x^{\alpha}} 
+ \lambda   (\Phi^2 - v^2)^2  \right).
\end{equation}
Here $\Phi$ is  a real scalar order parameter; $\lambda, \; v$ and $\gamma$ 
are positive constants; and $(x^{\alpha}),\; \alpha=1,2,3$, are Cartesian
coordinates in the usual $R^3$ space.  From (1) and (2) we obtain the 
following equation for the rescaled dimensionless order parameter $\phi =
\Phi/v$
\begin{equation}
\gamma \frac{\partial \phi}{\partial t} = \Delta \phi  - 2 \lambda v^2
\phi (\phi^2 -1),
\end{equation}
where $\Delta = \partial_{\alpha}\partial_{\alpha}$.
The domain wall solutions of Eq.(3) smoothly interpolate between $\phi = -1$
on one side of the wall and $\phi = +1$ on the other side.  Canonical
example of such a solution is given by the formula
\begin{equation}
\phi_0(x^3) = \tanh\frac{x^3 -a}{2l_0},
\end{equation}
where $l_0^{-2} = 4 \lambda v^2$ and $a$ is an arbitrary constant. The 
presence of the constant $a$ is due to translational invariance of Eq.(3).
This particular solution represents planar static domain wall located at 
the plane $x^3 =a$. Such domain wall is homogeneous along that plane.  
Its transverse profile is parametrized by $x^3$. Width of the wall is 
approximately equal to $l_0$, in the sense that for $|x^3 - a| \gg l_0$ 
values of $\phi$ differ from +1 or -1 by exponentially small terms. 
The form (4) of $\phi_0 (x^3)$ coincides with one-dimensional static 
kink present in one-dimensional version
of the model defined by (1) and (2). The domain wall can be regarded as 
embedding of that kink in the 3-dimensional space $R^3$. 

Notice that somewhere inside any domain wall there is a surface on which 
$\phi$ vanishes. For example, in the case of planar domain wall (4) 
$\phi_0=0$ for $x^3=a$. Such surface is called the core of the domain wall. 

The first step in our construction of the perturbative scheme consists in
introducing special coordinates comoving with the domain wall. One 
coordinate, let say $\xi$, parametrizes direction perpendicular to the 
domain wall, two other coordinates ($\sigma^1,\; \sigma^2$) 
parametrize the domain wall 
regarded as a surface in the $R^3$ space. The comoving coordinates have been 
proposed in the context of CEA method, \cite{5}. We introduce one 
important modification: an auxiliary surface $S$, which is present 
in the definition of the comoving coordinates, is apriori 
independent of the domain wall.  
In literature on CEA method it is defined directly 
in terms of the domain wall  --- the most popular choice is that $S$ 
coincides with the core. It turns out that the latter choice in general 
is not compatible with certain consistency conditions which appear in our
approximation scheme. Transformations to comoving coordinates in the 
cases of relativistic domain walls and vortices in Minkowski 
space-time can be found in \cite{9}, \cite{13}, respectively. Below we 
introduce such coordinates for the nonrelativistic domain wall moving in the 
$R^3$ space.

After all these remarks let us finally define the comoving
coordinates. We consider a smooth, closed or infinite surface $S$ in 
the usual $R^3$ space. It is close to the core and its shape roughly gives
the shape of the domain wall. In particular one may assume that
$S$ coincides with the core at certain time $t_0$. 
Points of $S$ are given by  $\vec{X}(\sigma^i,t)$, where $\sigma^i$ $(i=1,2)$
are two intrinsic coordinates on $S$, and $t$ denotes the time --- we
allow for motion of $S$ in the space. The vectors $\vec{X}_{,k},\; k=1,2$,
are tangent to $S$ at the point $\vec{X}(\sigma^i,t)$ \footnote{We use the 
compact notation $f_{,k} \equiv \partial f / \partial\sigma^k$.}. They are 
linearly independent, but not necessarily orthogonal to each other. At each 
point $\vec{X}(\sigma^i,t)$ of $S$ we also introduce a third vector 
$\vec{p}(\sigma^i,t)$ which is perpendicular to $S$, that is
\[ \vec{p} \vec{X}_{,k} =0. \]
We assume that $\vec{p}$ has unit length, $\vec{p}^2 =1$. The three vectors
$(\vec{X}_{,k}, \vec{p})$ form a local basis at the point 
$\vec{X}(\sigma^i,t)$ of $S$. With this basis given at each point of $S$, we
introduce geometric characteristics of $S$: 
induced metric tensor on $S$
\[ g_{ik} = \vec{X}_{,i} \vec{X}_{,k},   \]
and the extrinsic curvature coefficients 
\[ K_{il} = \vec{p}\vec{X}_{,il} \]
($i,k,l=1,2$).
They appear in Gauss-Weingarten formulas  
\begin{equation}
\vec{X}_{,ik} = K_{,ik} \vec{p} + \Gamma^l_{ik} \vec{X}_{,l}, \;\;\;\;
\vec{p}_{,k} = - g^{il} K_{lk} \vec{X}_{,i}. 
\end{equation}
Here the matrix $(g^{ik})$ is by definition the inverse of the matrix 
$(g_{kl})$, i.e.
$g^{ik}g_{kl}= \delta^i_l$, and $\Gamma^l_{ik}$ are Christoffel symbols
constructed from the metric $g_{ik}$. The two by two matrix $(K_{ik})$ is
symmetric. Two eigenvalues  $k_1,k_2$ of the matrix $(K^i_j)$, where 
$K^i_j = g^{il}K_{lj}$, are called extrinsic curvatures of $S$ at the 
point $\vec{X}$. The two main curvature radia are defined as $R_i =1/k_i$. 
In general they vary along $S$ and with time. 

The comoving coordinates $(\sigma^1, \sigma^2,\xi)$ at the time $t$ are
introduced by the following formula
\begin{equation}
\vec{x} = \vec{X}(\sigma^i,t) + \xi \vec{p}(\sigma^i,t).
\end{equation}
Here $\vec{x}=(x^{\alpha}), \alpha=1, 2, 3$,  
are the usual Cartesian coordinates
in the space $R^3$. $\xi$ is the coordinate in the direction perpendicular
to $S$. Notice that this direction has very simple parametrization ---
the r.h.s. of formula  (6) is a linear function of $\xi$. 
We will use a compact notation  for the comoving 
coordinates: $(\sigma^1, \sigma^2, \xi) = (\sigma^{\alpha})$,
with $\alpha$=1, 2, 3 and $\sigma^3 = \xi$. The
coordinates $(\sigma^{\alpha})$
are just a special case of 
curvilinear coordinates in $R^3$. The corresponding
metric tensor $G_{\alpha \beta}$ in $R^3$ has the following components:
\[ G_{33} =1, \;\; G_{3k} = G_{k3} =0, \;\; G_{ik} = N^l_i g_{lr}N^r_k, \]
where 
\[ N^l_i = \delta^l_i - \xi K^l_i, \]
$i,k,l,r =1,2$. Simple calculations give 
\[ \sqrt{G} =\sqrt{g} N, \]
where $G=det(G_{\alpha\beta}), 
\;\; g=det(g_{\alpha\beta})$, and $ N=det(N^i_k)$ is given by the following
formula 
\[ N = 1 - \xi K^i_i + \frac{1}{2} \xi^2 (K_i^i K^l_l - K^i_lK^l_i). \]
Components $G^{\alpha \beta}$ of the inverse metric tensor have the form  
\[ G^{33} =1,\;\; G^{3k}=G^{k3}=0, \;\; G^{ik}= (N^{-1})^i_r g^{rl} 
(N^{-1})^k_l, \]
where 
\[ (N^{-1})^i_r = \frac{1}{N} 
\left((1-\xi K^l_l) \delta^i_r + \xi K^i_r \right). \]
We see that dependence of $G_{\alpha \beta}$ on the transverse coordinate 
$\xi$ is explicit, and that $\sigma^1, \sigma^2$ appear through the tensors
$g_{ik}, \; K^l_r$ which characterise the surface $S$. 

In general the coordinates $(\sigma^{\alpha})$ have certain finite region of
validity. In particular, the range of $\xi$ is given by the smallest
positive $\xi_0(\sigma^i,t)$ for which $N =0$. It is clear that such $\xi_0$ 
increases with decreasing extrinsic curvature coefficients $K_i^l$, reaching
infinity for the planar domain wall.  We assume that the surface $S$ (and 
the domain wall) is smooth enough, so that outside of that region there are 
only exponentially small tails of the domain wall which give negligible 
contributions to physical characteristics of the domain wall. 

The comoving coordinates are utilised to write Eq.(3) in a form suitable 
for calculating the curvature corrections.  Laplacian $\Delta\phi$
in the new coordinates has the form 
\[ 
\Delta\phi = \frac{1}{\sqrt{G}} \frac{\partial}{\partial\sigma^{\alpha}}
\left(\sqrt{G} G^{\alpha\beta} \frac{\partial \phi}{\partial \sigma^{\beta}}
\right).
\] 
The time derivative on the l.h.s. of Eq.(3) is taken under the condition 
that all $x^{\alpha}$ are constant. It is convenient to use time derivative 
taken at constant $\sigma^{\alpha}$. The two derivatives are related by 
the formula
\[
\frac{\partial}{\partial t}|_{x^{\alpha}} = 
\frac{\partial}{\partial t}|_{\sigma^{\alpha}} +
\frac{\partial \sigma^{\beta}}{\partial t}|_{x^{\alpha}}
\frac{\partial}{\partial \sigma^{\beta}},
\]
where
\[ \frac{\partial \xi}{\partial t}|_{x^{\alpha}} 
= - \vec{p}\dot{\vec{X}}, \;\;\;
\frac{\partial \sigma^i}{\partial t}|_{x^{\alpha}} = - (N^{-1})^i_k
g^{kr} \vec{X}_{,r} (\dot{\vec{X}} + \xi \dot{\vec{p}}), 
\]
the dots stand for $\partial /\partial t |_{\sigma^i}$. 
The final step consists in rescaling the transverse variable $\xi$
\[ \xi = 2l_0 s. \] 
The dimensionless variable $s$ measures the distance from the surface $S$
in the unit $2l_0$. 
Equation (3) transformed to the comoving coordinates with $\xi$ rescaled
as above has the following form
\[
2 \gamma l_0^2 \left( \frac{\partial \phi}{\partial t}|_{\sigma^{\alpha}}
- \frac{1}{2l_0}\vec{p}\dot{\vec{X}} \frac{\partial\phi}{\partial s}
- (N^{-1})^i_k g^{kr} \vec{X}_{,r} (\dot{\vec{X}} + 2 l_0 s
\dot{\vec{p}}) \frac{\partial\phi}{\partial\sigma^i} \right)
\]
\begin{equation}
= \frac{1}{2} \frac{\partial^2 \phi}{\partial s^2} + \phi - \phi^3
+ \frac{1}{2N} \frac{\partial N}{\partial s} 
\frac{\partial \phi}{\partial s} + 2 l_0^2 \frac{1}{\sqrt{g} N} 
\frac{\partial}{\partial \sigma^j} \left(G^{jk}\sqrt{g} N 
\frac{\partial \phi}{\partial \sigma^k} \right), 
\end{equation}
which is convenient for construction of the expansion in  width.

\section{The expansion in the width}
We seek domain wall solutions of Eq.(7) in the form of expansion with 
respect to $l_0$, that is 
\begin{equation} 
\phi = \phi_0 + l_0 \phi_1 + l_0^2 \phi_2 + ...   .
\end{equation}
Inserting formula (8) in Eq.(7) and keeping only terms of the lowest
order ($\sim l_0^0$) we obtain the following equation 
\begin{equation}
\frac{1}{2} \frac{\partial^2 \phi_0}{\partial s^2} + \phi_0 - \phi_0^3
= 0.
\end{equation}
It has the well-known kink solutions 
\[
\phi_0 = \tanh(s-s_0),
\]
which formally have the same form as the planar domain walls (4). In the
remaining part of the paper we shall calculate curvature corrections to 
the simplest solution
\begin{equation}
\phi_0 = \tanh s.
\end{equation}
Notice that $\phi_0$ interpolates between the vacuum solutions $\pm 1$.
Therefore, the corrections $\phi_k, \; k \geq 1,$ should 
vanish in the limits $s \rightarrow \pm \infty$.  

Equations for the corrections $\phi_k, \; k\geq1,$  are obtained  by 
expanding the both sides of Eq.(7) and equating terms proportional
to $l_0^k$. They can be written in the form 
\begin{equation}
\hat{L} \phi_k = f_k,
\end{equation}
where 
\begin{equation}
\hat{L} = \frac{1}{2} \frac{\partial^2}{\partial s^2} + 1 - 3\phi_0^2 =
\frac{1}{2} \frac{\partial^2}{\partial s^2} + \frac{3}{\cosh^2 s} -2,
\end{equation}
and $f_k$ depends on the lower order contributions $\phi_l, \; l<k$.
Straightforward calculations give 
\begin{equation}
f_1 = \partial_s\phi_0 ( K^r_r - \gamma \vec{p}\dot{\vec{X}}),
\end{equation}
\begin{equation}
f_2 = 3 \phi_0 \phi_1^2 + 2s \partial_s\phi_0 K^i_j K^j_i +
\partial_s\phi_1 ( K^r_r - \gamma \vec{p}\dot{\vec{X}}),
\end{equation}
\begin{eqnarray}
& f_3 = 2 \gamma (\partial_t\phi_1 - g^{kr}\vec{X}_{,r}\dot{\vec{X}} 
\partial_k\phi_1 ) + 6 \phi_0\phi_1\phi_2 + \phi_1^3  & \nonumber \\
& + 2 s \partial_s\phi_1 K^i_jK^j_i - 2 s^2 \partial_s\phi_0 K^r_r 
\left( (K^i_i)^2 - 3 K^i_jK^j_i\right) & \nonumber \\
& - \frac{2}{\sqrt{g}}\partial_j(\sqrt{g}g^{jk}\partial_k\phi_1) +
\partial_s\phi_2 ( K^r_r - \gamma \vec{p}\dot{\vec{X}}), &
\end{eqnarray}
and
\begin{eqnarray}
& f_4 = 2\gamma ( \partial_t\phi_2 
- 2s g^{ik} \dot{\vec{p}}\vec{X}_{,k}\partial_i\phi_1 )
- 2 \gamma g^{jk} \vec{X}_{,k}\dot{\vec{X}} (\partial_j\phi_2 +
2s K^i_j\partial_i\phi_1 ) & \nonumber \\
& + 3 \phi_0 \phi_2^2 + 6 \phi_0 \phi_1 \phi_3 + 3 \phi_1^2 \phi_2 
+ 2 s \partial_s \phi_2 K^i_jK^j_i   & \nonumber \\
& - 4 s^3 \partial_s\phi_0 \left( (K^r_r)^4 - (K^r_sK^s_r)^2 
- 2 (K^r_r)^2 K^i_jK^j_i \right) & \nonumber \\
& - 2 s^2 \partial_s\phi_1 K^r_r \left(
(K^i_i)^2 - 3 K^i_jK^j_i \right) 
 - \frac{2}{\sqrt{g}}\partial_j(\sqrt{g}g^{jk}\partial_k\phi_2) & \nonumber \\
& -\frac{8s}{\sqrt{g}} \partial_j(\sqrt{g}K^{jk}\partial_k\phi_1)
+ 4s g^{jk}\partial_jK^r_r \partial_k\phi_1 +
\partial_s\phi_3 ( K^r_r - \gamma \vec{p}\dot{\vec{X}}), &
\end{eqnarray}
where $\partial_t = \partial /\partial t,\;\; \partial_i = \partial / 
\partial \sigma^i$.

Notice that all Eqs.(11) for $\phi_k$ are linear. The only nonlinear 
equation in our perturbative scheme is the zeroth order equation (9). 

Now comes the crucial point: operator $\hat{L}$ has a zero-mode, that is
a normalizable function $\psi_0(s)$ such that 
\[
\hat{L} \psi_0 =0.
\]
This function can be obtained by differentiating $\phi_0(x^3)$ given by 
formula (4) with respect to $a/2l_0$ and putting $a=0$,
\begin{equation}
\psi_0 = \frac{1}{\cosh^2 s}.
\end{equation}
This zero-mode owes its existence to invariance of Eq.(3) with respect to 
spatial translations, therefore it is often called the translational 
zero-mode. Let us multiply Eqs.(11) by $\psi_0(s)$ and integrate over $s$.
Because 
\[
\int^{\infty}_{-\infty} ds \psi_0 \hat{L} \phi_k = 0, 
\]
we obtain the consistency (or integrability) conditions 
\begin{equation}
\int^{\infty}_{-\infty} ds \psi_0(s) f_k(s) =0.
\end{equation}
The operator $\hat{L}$ appears also in the expansion in width for 
relativistic domain walls \cite{9}. Using standard methods \cite{15},\cite{9}
one can obtain the following formula for vanishing in the limits 
$s \rightarrow \pm \infty$ solutions $\phi_k$ of Eqs.(11):
\begin{equation}
\phi_k = G[f_k] + C_k(\sigma^i, \tau) \psi_0(s), 
\end{equation}
where
\begin{equation}
G[f_k] = - 2 \psi_0(s) \int^s_0 dx \psi_1(x) f_k(x) + 2 \psi_1(s) 
\int^s_{-\infty} dx \psi_0(x) f_k(x). 
\end{equation}
Here $\psi_0(s)$ is the zero-mode (17) and 
\begin{equation}
\psi_1(s) = \frac{1}{8} \sinh(2s) + \frac{3}{8} \tanh s + \frac{3}{8}
\frac{s}{\cosh^2s} 
\end{equation}
is the other solution of the homogeneous equation 
\[ 
\hat{L} \psi = 0.
\]
The second term on the r.h.s. of formula (19) obeys the homogeneous 
equation $\hat{L}\phi_k =0$. It vanishes when $s \rightarrow \pm \infty$. 

One can worry that $\phi_k, \; k\geq 1$, given by formulas (19), (20) 
do not vanish when $s\rightarrow \pm \infty$ because the second term on 
the r.h.s. of formula (20) is proportional to $\psi_1$ which 
exponentially increases in the limits $s \rightarrow \pm \infty$. 
However, the integrals 
\[ \int^s_{-\infty} dx \psi_0 f_k \]
vanish in that limit, see the consistency conditions (18). Moreover, 
qualitative analysis of Eq.(7) shows  that $f_k \sim 
(\mbox{polynomial in}\; s) \times \exp(-2|s|)$ for large $|s|$, hence those
integrals behave like $(\mbox{polynomial in}\;s) \times \exp(-4|s|)$ for 
large $|s|$ ensuring that $\phi_k$ exponentially  vanish when 
$|s| \rightarrow \infty$.

We have explicitly solved Eqs.(11). The solutions (19) contain as yet 
arbitrary functions $C_k(\sigma^i, t)$, and also $\vec{X}(\sigma^i, t)$
giving points of the comoving surface $S$ --- $K^i_l,\; g_{ik}$ follow from 
$\vec{X}$. It turns out that the conditions 
(18) are so restrictive that they essentially fix these functions. The
consistency condition with  $k=1$ is equivalent to
\begin{equation}
\gamma \vec{p} \dot{\vec{X}} = K^r_r,
\end{equation}
where we have used formulas (10), (13) and (17). 
Thus we have obtained equation 
for $\vec{X}$. It is of the same type as Allen-Cahn equation \cite{16}, 
but in our approach it describes motion of the auxiliary surface $S$ only.
Equation (22) should be compared with Nambu-Goto equation for a 
relativistic membrane obtained in the 
relativistically invariant version of our model \cite{9}. We shall see 
below that  the remaining consistency  conditions do not give more 
restrictions for $\vec{X}$ at least  up to the  fourth order --- they can be 
saturated by the functions $C_k(\sigma^i, t)$. We expect that this is true
to all orders but we have not attempted to provide a proof. 

Let us now proceed with the discussion of the perturbative corrections:
Taking into account the condition  (22) we have $f_1 =0$. Therefore
\begin{equation}
\phi_1 = \frac{C_1(\sigma^i,t)}{\cosh^2s}.
\end{equation}
Equation (11) with $k=1$ does not provide more information. 

The second order contribution $\phi_2$ is given by formula (19) with $k=2$.
Using the results (22, 23) from the first order we obtain the following
formula
\begin{equation}
\phi_2 = \psi_2(s) C_1^2(\sigma^i,t) + \psi_3(s) K^i_jK^j_i + 
\frac{C_2(\sigma^i,t)}{\cosh^2s},
\end{equation}
where 
\[ \psi_2(s) = - \frac{\sinh s}{\cosh^3s}, \]
\[ \psi_3(s) = \frac{-4}{\cosh^2s} \int^s_0dx  \frac{ x \psi_1(x)}{\cosh^2x} 
+ 4 \psi_1(s) \int^s_{-\infty} dx \frac{x}{\cosh^4x}.    \]
$\psi_3(s)$ can easily be evaluated, e. g., numerically. Also the higher order
corrections involve only rather simple integrals of elementary functions.

The consistency condition (18) with $k=2$ does not give any restrictions ---
it can be reduced to the identity  $0 = 0$. More interesting is the next 
condition, that is the one with $k=3$. It can be written in the form of 
the following inhomogeneous equation for $C_1(\sigma^i,t)$ 
\begin{eqnarray}
& \gamma ( \partial_t C_1 - g^{kr} \vec{X}_{,r}\dot{\vec{X}}
\partial_k C_1 )  - \frac{1}{\sqrt{g}}
\partial_j(\sqrt{g} g^{jk}\partial_kC_1)
- K^i_j K^j_i C_1 & \nonumber \\
& = \frac{1}{2} (\frac{\pi^2}{6}-1)
K^r_r\left((K^i_i)^2 - 3 K^i_jK^j_i\right). &
\end{eqnarray}
This equation determines $C_1$ provided that we fix initial data for it.
The consistency condition coming from the fourth order ($k=4$) is 
equivalent to the following homogeneous equation for $C_2$
\begin{equation}
\gamma ( \partial_t C_2 - g^{kr} \vec{X}_{,r}\dot{\vec{X}}
\partial_k C_2 ) - 
\frac{1}{\sqrt{g}} \partial_j(\sqrt{g} g^{jk}\partial_kC_2)
-  K^i_j K^j_i C_2  = 0. 
\end{equation}

The perturbative scheme presented above is not quite straightforward.
Therefore  we would like to add several explanations. The formulas presented
above give a whole family of domain wall solutions. 
To obtain one concrete domain wall solution we have to choose initial
position of the auxiliary surface $S$. Its positions at later times are 
determined from Eq.(22). We also have to fix initial values of the 
functions $C_1, C_2$ and to find the corresponding solutions of Eqs. (25), 
(26).  The approximate domain wall solution $\phi$ is given by formulas
(8), (19), (20).  Notice that we are not allowed to choose the 
initial profile of the domain wall arbitrarily because the dependence 
on the transverse coordinate $s$ is explicitly given by these formulas.
Any choice of the initial data gives an approximate domain wall solution. 
Of course such a choice should not lead to large perturbative corrections 
at least in certain finite time interval. Therefore one should require that
at the initial time $l_0 C_1 \ll 1 , l_0^2 C_2 \ll 1, l_0 K^i_j \ll 1$. 
The domain wall is located close to the surface $S$ because for large 
$|s|$ the perturbative contributions vanish and the leading term $\tanh s$ 
is close to one of the vacuum values $\pm1$.

The presented formalism is invariant with respect to changes of coordinates
$\sigma^1, \sigma^2$ on $S$. In particular, in a vicinity of any point 
$\vec{X}$ of $S$ we can choose the coordinates in such a way that 
$g_{ik} =\delta_{ik}$ at $\vec{X}$. In these coordinates Eq.(22) has the 
form 
\begin{equation}
\gamma v = \frac{1}{R_1} + \frac{1}{R_2},
\end{equation}
where $v$ is the velocity in the direction $\vec{p}$ perpendicular to $S$ and 
$R_1, R_2$ are the main curvature radia of $S$ at the point $\vec{X}$. 

Let us present a simple example: take $S$ to be a sphere of radius $R$. 
Then $R_1 =R_2 = - R(t)$, $v=\dot{R}$ and Eq.(27) gives 
\[ R(t) = \sqrt{R^2_0 - \frac{4}{\gamma}(t-t_0)},    \]
where $R_0$ is the initial radius.  Our approximate formulas are expected
to be meaningful as long as $R(t)/l_0 \gg 1$. Equations (25), (26) 
reduce to
\[ 
\gamma \partial_t C_1  - \frac{1}{R^2}\left(\frac{1}{\sin\theta}
\partial_{\theta}(\sin\theta \partial_{\theta}C_1) + \frac{1}{\sin^2\theta}
\partial^2_{\phi}C_1 \right)- \frac{2}{R^2} C_1 = 2 (\frac{\pi^2}{6} -1) 
\frac{1}{R^3},  
\]
\[
 \gamma \partial_t C_2 - \frac{1}{R^2}\left(\frac{1}{\sin\theta}
\partial_{\theta}(\sin\theta \partial_{\theta}C_2) + \frac{1}{\sin^2\theta}
\partial^2_{\phi}C_2 \right) - \frac{2}{R^2} C_2  = 0. 
\]
In the last equation $\theta, \phi$ are the usual spherical coordinates on
$S$ (we apologize for using letter $\phi$ also in this meaning). If we take 
the simplest initial data at $t=t_0$, namely $C_1 = C_2 = 0$, then the last 
equation implies that $C_2 = 0$ also for $t>t_0$ while $C_1 > 0$.
The Cartesian coordinate frame is located at the center of the sphere $S$
and $\vec{p}$ is the outward normal to $S$; $\;$ $s = (r-R(t))/2l_0$, where 
$r$ is the radial coordinate in $R^3$.

\section{Remarks}
We would like to make several general remarks about the expansion in width. 

\noindent 1. We have used $l_0$ as a formal expansion parameter. It is a
dimensionful quantity, hence it is hard to say whether its value is small 
or large. 
What really matters is smallness of the corrections $l_0 \phi_1,\; 
l_0^2 \phi_2$. This is the case if $l_0 C_1 \ll 1, \; l_0^2 C_2 \ll 1$
and $l_0 K^i_j \ll 1$, as seen from formulas (8), (19) and (20).

\noindent 2. Notice that an assumption that $S$ coincides with the core
for all times in general would not be compatible with the expansion in width. 
If we assume that $C_1 = 0 = C_2$ at certain initial time $t_0$, Eq.(25) 
implies that $C_1 \neq 0$ at later times (unless the r.h.s. of it happens
to vanish). Then, it follows from formulas (8), (10) and (23) that
$\phi$ does not vanish at $s=0$, that is on $S$.

\noindent 3. The question of convergence of the expansion (8) has not been
analysed. Actually we think that the expansion can turn out to be convergent,
in spite of the fact that more frequent in field theory are asymptotic
expansions. Moreover, this problem seems to be within the reach of the 
present day mathematical  techniques. 

\noindent 4. Finally, we would like to stress that we have abandoned effects
which come from perturbations of the exponential tails of the domain wall. 
For example, if we have a domain wall in the form of cylinder with very 
large radius and small height (and with rounded edges), then the top and 
bottom parts are flat, and according to Eq.(27) they do not move. In our 
approximation the
cylinder shrinks from the sides where the mean curvature $1/R_1 + 1/R_2$
does not vanish. Now, in reality the top and bottom parts interact with 
each other. This interaction is very small only if the two flat parts are 
far from each other. We have neglected it altogether assuming the $\tanh s$
asymptotics for large $s$. Thus, our approximate solution takes into account
only the effects of curvature.

\section{Acknowledgement}
I would like to thank the organizers of the School for the possibility to
present this work, and for stimulating atmosphere during the School.  \\
This paper is supported in part by KBN grant 2P03B 095 13.

\end{document}